\documentclass[pra,twocolumn,showpacs]{revtex4}
\usepackage{dcolumn}
\usepackage{graphicx}
\newcommand{\nn}{\nonumber}
\usepackage{bm}

\begin{document}

\title{Coreless and singular vortex lattices in rotating spinor Bose-Einstein condensates}

\author{Takeshi Mizushima}
\email{mizushima@mp.okayama-u.ac.jp}
\affiliation{Department of Physics, Okayama University, Okayama 700-8530, Japan}
\author{Naoko Kobayashi}
\affiliation{Department of Physics, Okayama University, Okayama 700-8530, Japan}
\author{Kazushige Machida}
\affiliation{Department of Physics, Okayama University, Okayama 700-8530, Japan}
\date{\today}

\begin{abstract}
We theoretically investigate vortex-lattice phases of rotating spinor Bose-Einstein condensates (BEC) with the ferromagnetic spin-interaction by numerically solving the Gross-Pitaevskii equation. The spinor BEC under slow rotation can sustain a rich variety of exotic vortices due to the multi-component order parameters, such as the Mermin-Ho and Anderson-Toulouse coreless vortices (the 2-dimensional skyrmion and meron) and the non-axisymmetric vortices with the sifting vortex cores. Here, we present the spin texture of various vortex-lattice states at higher rotation rates and in the presence of the external magnetic field. In addition, the vortex phase diagram is constructed in the plane by the total magnetization $M$ and the external rotation frequency $\Omega$
by comparing the free energies of possible vortices. It is shown that the vortex phase diagram in a $M$-$\Omega$ plane may be divided into two categories; (i) the coreless vortex lattice formed by the several types of Mermin-Ho vortices and (ii) the vortex lattice filling in the cores with the pure polar (antiferromagnetic) state. In particular, it is found that the type-(ii) state forms the composite lattices of coreless and polar-core vortices. The difference between the type-(i) and type-(ii) results from the existence of the singularity of the spin textures, which may be experimentally confirmed by the spin imaging  within polarized light recently proposed by Carusotto and Mueller. We also discussed on the stability of triangular and square lattice states for rapidly rotating condensates.
\end{abstract}


\maketitle

\section{Introduction}

The quantized vortex is one of the hallmark of superfluidity and such the macroscopic quantum
phenomenon has been studied in many different physical fields ranging 
from the condensed matter physics to neutron stars and cosmology.
Recently, quantized vortices have been successfully created in Bose-Einstein condensates (BEC's) 
of alkali-atom gases confining in a magnetic trap by several experimental methods \cite{nist,ens,jila,leanhardt}.
The static and dynamic properties of vortices and vortex-lattices 
have been investigated both theoretically and experimentally \cite{vortex}.
Several groups are now able to prepare a vortex array with vortices 
more than one hundred in a BEC \cite{abo,coddington}.

Further achievement of Bose-Einstein condensation was recently done 
by all optical methods without recourse to the magnetic trapping 
in $^{23}$Na \cite{stenger,gorlitz}, $^{87}$Rb \cite{barrett,chang,schmaljohann}, 
and $^{174}$Yb atoms \cite{takasu}.
Apart from $^{174}$Yb atoms, these systems, so-called spinor BEC's,
can keep internal degrees of freedom with the hyperfine spin $F\!=\!1$ or 2.
The two-body interaction of spin-1 bosons are written as 
$g_n + g_s \mbox{\boldmath $F$}\cdot \mbox{\boldmath $F$}$ 
with two parameters $g_n$ and $g_s$ which describe density 
and spin interactions, respectively \cite{ohmi,ho}.
The spin interaction of $^{23}$Na spinor condensates is antiferromagnetic $g_s>0$ \cite{stenger},
while it is shown by Klausen {\it et al.} \cite{klausen}
that $^{87}$Rb condensates have a ferromagnetic nature $g_s<0$.
Thus, we now have spinor BEC's with both ferromagnetic and antiferromagnetic interactions.

Due to this internal degree of freedom, a rich variety of exotic topological excitations
have been proposed by a large number of authors 
after the pioneering work by Ohmi and Machida \cite{ohmi} and Ho \cite{ho}: 
monopoles \cite{stoof,martikainenPRL,savage}, 
3-dimensional skyrmions \cite{khawaja} 
and 2-dimensional skyrmions (coreless vortices) \cite{marzlin,tuchiya,mizushimaPRL,reijndersPRA},
alice strings (half-quantum vortices) \cite{ruostekoski,leonhardt,zhou}, 
and the other unconventional vortices 
\cite{yip,isoshimaJPS,isoshimaPRA,mizushimaPRA,martikainen,kita,mueller,reijndersPRL}. 
Spinor BEC's provide us an opportunity to minutely study these exotic properties as a new example of 
multi-component superfluids.
Among such exotic state, in particular, skyrmion excitations play an important role 
in the other physics systems: the quantum field theory \cite{rajaraman}, 
the quantum Hall system \cite{sondhi}, superfluid $^3$He \cite{salomaa}, 
nematic liquid crystals \cite{bogdanov}, and unconventional superconductivity \cite{morinari}.
The periodic structures of such the topological state have been studied 
in the rotating superfluid $^3$He \cite{fujita,fetter,salomaa,karimaki,kitaPRL2001}, 
called the Mermin-Ho \cite{mermin} and Anderson-Toulouse vortices \cite{anderson}.
In the quantum Hall system, the textures and stability of skyrmion lattices 
have been also studied \cite{brey,green}. 

The earlier theoretical study focused on skyrmion excitations in a spinor BEC
has been performed by Khawaja and Stoof \cite{khawaja}, who pointed out 
that the 3-dimensional skyrmion is not a thermodynamically stable object.
The numerical analysis of the Gross-Pitaevskii and Bogoliubov equations under slow rotation,
however, led to the conclusion that such the topological excitation in the 2-dimensional disk 
is stable and robust in the ferromagnetic spin-interaction \cite{mizushimaPRL}. 
In addition, at the high rotation limit, Reijnders {\it et al.} \cite{reijndersPRL} 
have presented vortex- and skyrmion-lattice states  
for the spin-1 bosons in the lowest Landau level.
They have also analyzed the exact ground state near the critical rotation
and have proposed a rich variety of the quantum Hall liquid state.
In the mean-field regime for rapidly rotating spinor BEC's, 
the numerical study of the Ginzburg-Landau equation has been performed 
within the higher Landau levels \cite{kita}. 
Here, it has been found that the several types of the vortex lattice
with the shift-core compete with each other, which sensitively depends on the spin interaction. 

Previously in our series of papers \cite{mizushimaPRL,mizushimaPRA},
we have identified that in spinor BEC's with the ferromagnetic spin-interaction, 
the Mermin-Ho vortex is favored under slow rotation. 
This coreless vortex naturally connects to the non-vortex state as the total magnetization becomes
higher, i.e., the scalar condensate limit.
Upon increasing the rotation rate, the stable region of the Mermin-Ho vortex 
shifts toward the low magnetization region,
while the another vortex state, called the polar-core vortex,
appears in the high magnetization region. 
This is the vortex state filling the core with the pure polar (antiferromagnetic) state. 
In this paper, we will investigate the structures of many vortices 
by numerically solving the Gross-Pitaevskii equation. 
The main purpose of the present paper is to address what is the sequence of states 
in a spinor BEC with the higher rotation frequency and complete the vortex phase diagram in the plane;
the external rotation frequency $\Omega$ versus the magnetization $M$ in order to 
help establishing the properties of textures in spinor BEC's.
In addition, we will discuss on the stability of two lattice states for rapidly rotating condensates,
such as the Abrikosov lattice filling the cores with the polar state 
and the square lattice having the continuous texture \cite{fujita}.

This paper is organized as follows.
In Sec. II, we first present the extended Gross-Pitaevskii equation for spinor BEC's, 
and then explain the numerical procedure to find local minima of the energy functional.
The spin textures and other properties of the favored single-vortex state is shown in Sec. III. 
In Sec. IV, we present the vortex phase diagram in the $\Omega$-$M$ plane, 
obtained by comparing free energies. We also displayed the detailed structures of each ground state.
Furthermore, in Sec. V, 
we show the instability of triangular or square lattice states for rapidly rotating condensates
upon changing the total magnetization (or the external magnetic field).
The conclusion and discussions are given in Sec. VI.

\section{Theoretical Formulation}

\subsection{Gross-Pitaevskii equation}

The Hamiltonian for the $F=1$ spinor BEC in a frame rotating with the frequency 
$\mbox{\boldmath $\Omega$} = \Omega \hat{{\mbox{\boldmath $z$}}}$ is \cite{ohmi, ho}
\begin{eqnarray}
\mathcal{H} 
&=& \int d{\bf r} 
   \left[ \sum _{j}
       \Psi _j^{\dagger} \{h({\bf r}) - \mu - j B_z \} \Psi _j \right. \nn \\
&&   \left. 
      + \frac{1}{2} \sum _{ijkl}
         \Psi _i^{\dagger} \Psi _j^{\dagger} 
		 \left\{ g_n \delta _{jk} \delta _{il}
            + g_{\rm s} \mbox{\boldmath $F$} _{ik} \cdot \mbox{\boldmath $F$}_{jl} 
	     \right\}\Psi _k \Psi _l
\right],
\label{eq:H}
\end{eqnarray}
where $\Psi _{j}$ and $\Psi^{\dag}_j$ are the field creation and annihilation operators for a boson 
in the eigenstates of $F_z$  ($i,j,k,l = 0, \pm 1$). 
The one-body Hamiltonian is written as
\begin{eqnarray}
h({\bf r}) = - \frac{\hbar^2 \nabla^2}{2m}  + V({\bf r}) - \Omega L_z,
\end{eqnarray}
with the two-dimensional confinement potential 
$V({\bf r})\!=\!\frac{1}{2} m \omega^2 (x^2 + y^2)$ and 
the projection of the angular momentum to the $z$-axis 
$L_z\!=\!-i\hbar (x\partial _y - y \partial _x)$.
The interaction between atoms with the mass $m$ is characterized 
by the interaction strengths through the 
``density'' channel, $g_{n} = \frac{4 \pi \hbar^2}{m} \cdot \frac{a_0 + 2a_2}{3}$, 
and the ``spin'' channel, $g_{s} = \frac{4 \pi \hbar^2}{m} \cdot \frac{a_2 -  a_0}{3}$,
where $a_{0}$ and $a_{2}$ are the $s$-wave scattering length
with the total spin 0 and 2 channels, respectively.
$\mu$ and $B_z$ correspond to 
the chemical potential and magnetic field along the $z$-axis, respectively.

The spin angular momentum operators $F_{\alpha}$ ($\alpha\!=\!x,y,z$) with $F\!=\!1$  
can be  expressed in matrices as 
\begin{eqnarray}
&&F_x = \frac{1}{\sqrt{2}}
        \left( 
           \begin{array}{ccc}
               0 & 1 & 0\\
               1 & 0 & 1\\
               0 & 1 & 0
           \end{array}
        \right), \nn \\
&&F_y = \frac{i}{\sqrt{2}}
      \left(
            \begin{array}{ccc}
            0 & -1 &  0\\
            1 &  0 & -1\\
            0 &  1 &  0
            \end{array}
      \right), \\
&&F_z =
      \left( 
            \begin{array}{ccc}
               1 &  0 &  0\\
               0 &  0 &  0\\
               0 &  0 & -1
            \end{array}
      \right), \nn 
\end{eqnarray}
where the basis is taken as the eigenvector of the spin projection along $z$-axis. 
These operators satisfy the commutation relation, 
$[F_{\alpha}, F_{\beta}] = i \epsilon _{\alpha\beta\gamma}F_{\gamma}$.
In this basis, 
the field operators for spin-1 bosons are described as 
$\mbox{\boldmath $\Psi$} = (\Psi _{+1}, \Psi _{0}, \Psi _{-1})$.

Replacing $\mbox{\boldmath $\Psi$}$ in Eq.~(\ref{eq:H}) by the condensate wavefunction
$\mbox{\boldmath $\psi$}=\langle\mbox{\boldmath $\Psi$}\rangle$ and following 
the standard procedure, 
the time-dependent Gross-Pitaevskii (GP) equation is obtained as 
\begin{eqnarray}
i\hbar \frac{\partial}{\partial t}\psi _j =
    \left[
      \left\{ h - \mu -j B_z + g_{\rm n} \rho \right\} \delta_{jk} 
      + g_{\rm s} \langle \mbox{\boldmath $F$} \rangle \cdot \mbox{\boldmath $F$}_{jk} 
    \right] \psi_k, 
\label{eq:gp}
\end{eqnarray}
where $\langle A \rangle = \sum _{ij} \psi^{\ast}_i ({\bf r}) A_{ij} \psi_j ({\bf r})$ 
and the local density 
$\rho({\bf r}) \!=\! \mbox{\boldmath $\psi$}^{\dag}({\bf r})\cdot\mbox{\boldmath $\psi$}({\bf r})$.

Here, equilibrium states are found numerically via imaginary time propagation
of Eq.~(\ref{eq:gp}); $t \rightarrow \tau\!=\!-it$, starting from arbitrary initial state 
with random phases of each component and with random vortex configurations.
This numerical procedure is equivalent to find the local minima of the free energy functional
\begin{eqnarray}
F[\psi _j, \psi^{\ast}_j] 
&=& \ \int d{\bf r} \left\{ \langle h \rangle
 + \frac{1}{2} ( g_n \rho^2 + g_s \langle \mbox{\boldmath $F$} \rangle ^2 ) \right\} \nn \\
 &&     - \mu N_{\rm 2D} - B_z M ,
 \label{eq:free}
\end{eqnarray}
where $\mu$ and $B_z$ are interpreted as the Lagrange multipliers. 
We use the total number $N_{\rm 2D}$ and the total magnetization 
$M\!=\!\sum _j \int d{\bf r} j |\psi _j|^2$ as independent variables.
Since we assume uniformity along the $z$ direction, 
the order parameter must fulfill the normalization condition
\begin{eqnarray}
N_{\rm 2D} = \sum _j \int d{\bf r} |\psi _j ({\bf r})|^2.
\end{eqnarray}
In order to satisfy this condition, the chemical potential $\mu$ is varied during the numerical iteration.
The propagation in the imaginary time continues
until the fluctuation in $\mu$ becomes smaller than $10^{-10}$ and also that in the norm
becomes $\le 10^{-8}$

The actual calculations are carried out by discretizing the two-dimensional
space into $100^2$-$400^2$ mesh.
We have performed extensive search to find stable vortices, starting with arbitrary initial
vortex configurations for the ferromagnetic interaction strength, $g_s/g_n= -0.02$.
In addition, we use the following parameters:
the mass of a $^{87}$Rb atom $m$=1.44$\times 10^{-25}$kg,
the trapping frequency $\omega/2\pi$=200Hz,
and the particle number per unit length along the $z$ axis $N_{\rm 2D} \!=\! 10^4 / \mu$m.

\subsection{Local spin and nematic orders}

It is convenient to introduce a new set of basis, $\psi _{\alpha}$ ($\alpha \!=\! x, y, z$)
where the quantization axis is taken along the $\alpha$ direction \cite{ohmi}.
The transformation from the Cartesian representation $\psi _{\alpha}$ to $\psi _j$ is obtained 
as
\begin{eqnarray}
\mbox{\boldmath $\psi$} =
\left( 
      \begin{array}{c} 
         \psi_{+1} \\ \psi_{0}  \\ \psi_{-1}
      \end{array}
\right) 
= \left( 
        \begin{array}{ccc}
           \frac{-1}{\sqrt{2}} &    \frac{i}{\sqrt{2}}     &  0\\
           0 &    0     & 1 \\
           \frac{1}{\sqrt{2}} & \frac{i}{\sqrt{2}} &  0
        \end{array}
  \right)
  \left( 
        \begin{array}{c}
           \psi_{x}  \\ \psi_{y}  \\ \psi_{z}
        \end{array}
  \right)  .
\end{eqnarray}
Then it is useful to adopt the following representation with real vectors 
$\mbox{\boldmath $m$}$ and $\mbox{\boldmath $n$}$: 
\begin{eqnarray}
\mbox{\boldmath $\psi$}  = \sqrt{\rho}
   (\mbox{\boldmath $m$} + i\mbox{\boldmath $n$}).
\end{eqnarray}
We also define the {\it spin texture} as 
\begin{eqnarray}
\mbox{\boldmath $l$}
   = \mbox{\boldmath $m$} \times \mbox{\boldmath $n$},
\end{eqnarray}
which describes the direction of the local spin, 
$\mbox{\boldmath $l$} \!=\! \langle \mbox{\boldmath $F$} \rangle / \rho$.
Recently, Mueller \cite{mueller} shows that a tensor
$Q_{\alpha\beta}\!=\!\psi^{\ast}_{\alpha}({\bf r})\psi_{\beta}({\bf r})$  
can be decomposed as follows:
\begin{eqnarray}
&&Q_{\alpha\beta} ({\bf r}) 
  \!=\!   i \epsilon_{\alpha\beta\gamma} l_{\gamma} ({\bf r}) \rho({\bf r})
        + N_{\alpha\beta} ({\bf r}), \\
&&N_{\alpha\beta} ({\bf r}) 
  \!=\!   \delta_{\alpha\beta} \rho ({\bf r}) 
        - \frac{1}{2}\left[  \langle F_{\alpha}F_{\beta} \rangle
                           + \langle F_{\beta}F_{\alpha} \rangle \right],
\end{eqnarray}
where $N_{\alpha\beta} ({\bf r})$ is a symmetric tensor with second-rank and
describes the spin fluctuation. 
Furthermore, Carusotto and Mueller \cite{carusotto} propose that
the local values of these spin and nematic orders can be imaged by using the polarized light. 
Thus it may be convenient to define the {\it local nematicity} as
\begin{eqnarray}
{\mathcal N} = {\rm Tr} [\hat{N}^2]/\rho.
\end{eqnarray}
The amplitude of this nematicity is obtained as
\begin{eqnarray}
{\mathcal N} ({\bf r}) = 1 - \frac{1}{2} \sum _{\alpha} l^2_{\alpha} ({\bf r}) , 
\end{eqnarray}
which reflects the competing characteristics between the local spin order and  the local nematic order.

\section{Single-vortex states}

In the axisymmetric system, the order parameter is obtained within the 
winding number of the $j$-th component $w_j$ and the relative phase $\alpha_j$
as $\psi _j = |\psi_j| \exp{[i(w_j\theta+\alpha_j)]}$.
From the minimizing of the spin-interaction energy, $g_s \langle \mbox{\boldmath $F$} \rangle^2$,
it follows that the winding and the relative phase satisfy the relation 
$2w _{0} = w _{+1} + w _{-1}$
and $2\alpha _{0} = \alpha _{+1} + \alpha _{-1} + n\pi$ \cite{isoshimaJPS}, where $n$ is an integer and 
the even (odd) number corresponds to the ferromagnetic (antiferromagnetic) interaction.
The spinor order parameter then may be written as 
\begin{eqnarray}
\psi _j = \sqrt{\rho _j} \exp{[i(w\theta - j(w'\theta +\alpha))]},
\end{eqnarray}
and the spin texture in the $x$-$y$ plane is given as 
$(l_x, l_y)\propto(\cos{(w'\theta + \alpha)}, \sin{(w'\theta + \alpha)})$,
which is classified with the winding number $w'$ and the relative phase $\alpha$.
Here, $w$ and $w'$ are related to $w_j$ by $w_j = w - jw'$,
and $\alpha$ and $\alpha _j$ also satisfies the relation $\alpha _j = -j \alpha$.

\subsection{Coreless vortices}

\begin{figure}[b]
\includegraphics[width=8.0cm]{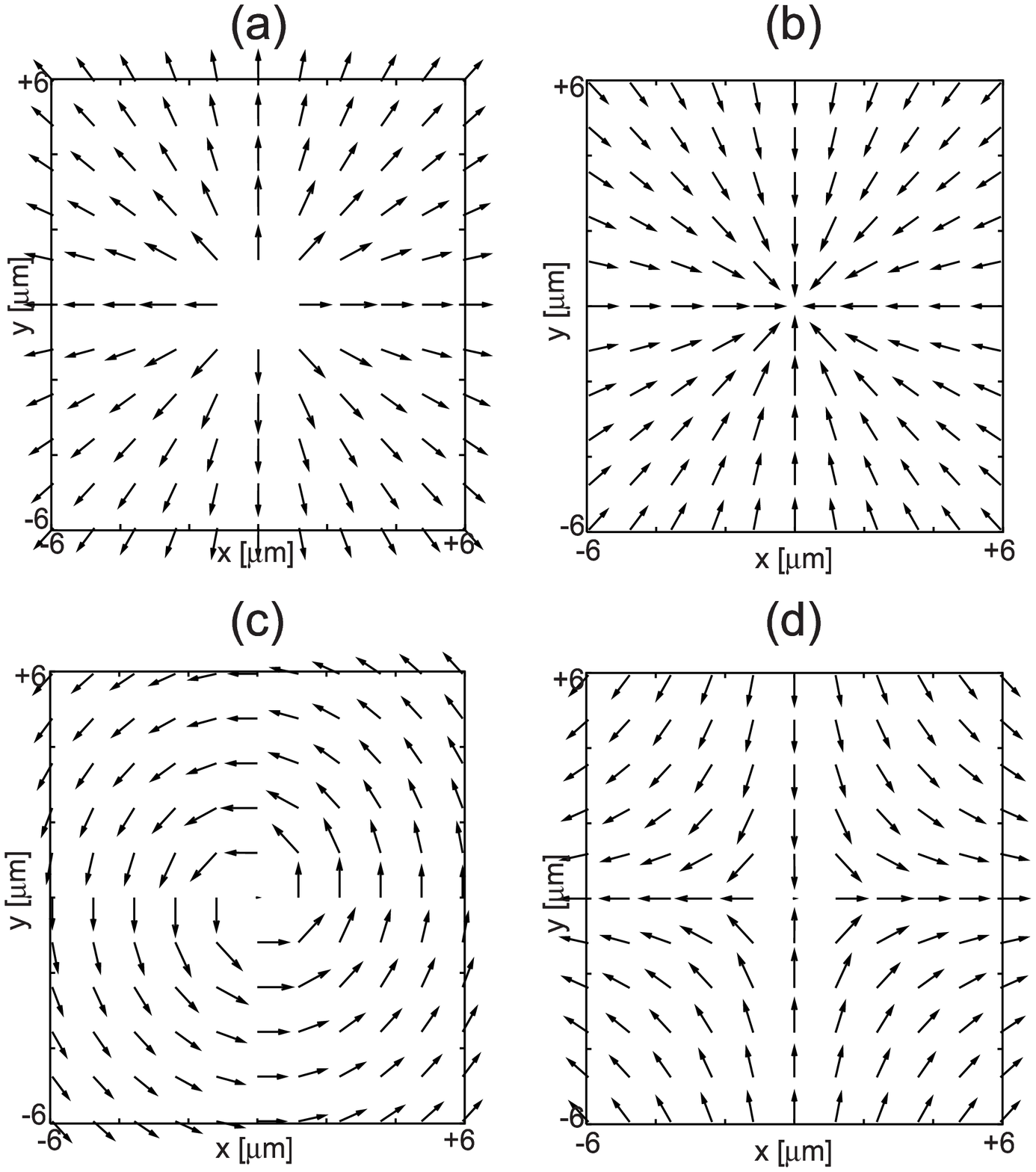}
\caption{The spin textures of isolated Mermin-Ho vortices;
(a) and (b) radial disgyrations ($w' \!=\! +1$, $\alpha \!=\! 0, \pi$). 
(c) the circular disgyration ($w'\!=\! +1$, $\alpha \!=\! \pi/2$). 
(d) the cross disgyration with $w ' = -1$. 
}
\label{fig:2dMH}
\end{figure}

In the ferromagnetic gas with $\mbox{\boldmath $l$}^2 \!=\! 1$,
$\mbox{\boldmath $m$}$ and $\mbox{\boldmath $n$}$ in Eq.~(\ref{eq:free})
are unit vectors with $\mbox{\boldmath $m$} \perp \mbox{\boldmath $n$}$
and hence, these three real vectors 
($\mbox{\boldmath $l$}$,$\mbox{\boldmath $m$}$,$\mbox{\boldmath $n$}$) 
form a triad \cite{ohmi}.
From the superfluid velocity 
$\mbox{\boldmath $v$} 
    = \sum_{\alpha} \frac{\hbar\rho}{2m}
      [m_{\alpha} \nabla n_{\alpha} - n_{\alpha} \nabla m_{\alpha}]$, 
the {\it local vorticity} is obtained as
\begin{eqnarray}
\nabla \times \mbox{\boldmath $v$} 
= \frac{\hbar}{m}\sum_{\alpha,\beta,\gamma} \epsilon_{\alpha\beta\gamma} 
  l_{\alpha} (\nabla l_{\beta}) \times (\nabla l_{\gamma}),
\end{eqnarray}
which implies the Mermin-Ho relation \cite{mermin}.
In other words, since the original Hamiltonian for the non-rotating system 
with the ferromagnetic spin-interaction 
has the $SO(3)$ symmetry \cite{ho}, the local spins may sweep the whole or half unit sphere.
The Mermin-Ho (MH) vortex is thermodynamically favored under slow rotation
and the weak magnetic field where the spin interaction 
rather than the contribution of the magnetic field dominates the free energy, Eq.~(\ref{eq:free}).
This coreless vortex takes $w \!=\! 1$ and $w'\!=\!1$, corresponding to the winding combination
$\langle w _{+1}, w _{0}, w _{-1} \rangle \!=\! \langle 0, 1, 2 \rangle$.
This is parametrized as 
\begin{eqnarray}
\mbox{\boldmath $\psi$} = \sqrt{\rho} e^{iw\theta}
\left(
\begin{array}{c}
e^{-i (w' \theta + \alpha)}\cos^2{\frac{\beta}{2}} \\
\sqrt{2} \sin{\frac{\beta}{2}} \cos{\frac{\beta}{2}} \\
e^{i(w' \theta + \alpha)} \sin^2{\frac{\beta}{2}} 
\end{array}
\right),
\end{eqnarray}
where the bending angle $\beta(r)$ runs over 
$0\leq\beta(r)\leq\pi$ and $\theta$ signifies the polar angle in polar coordinates. 
The spin texture is given as 
\begin{eqnarray}
\mbox{\boldmath $l$}({\bf r}) =
\left( 
\begin{array}{c}
   \sin{\beta(r)} \cos{(w'\theta+\alpha)} \\ \sin{\beta(r)}\sin{(w'\theta+\alpha)} \\ 
   \cos{\beta(r)}
\end{array}
\right),
\label{eq:spin}
\end{eqnarray}
where $\beta$ has the flexibility for $M$ \cite{mizushimaPRL}.
As shown in Figs.~{\ref{fig:2dMH}}(a), (b), and (c), 
the MH vortex has a rich variety of 
the two-dimensional type of spin textures by changing the relative phases 
in between spin components: Typical textures form
the radial disgyration $\mbox{\boldmath $l$} \parallel {\bf r}$ 
($\alpha\!=\!0, \pi$)  and the circular disgyration $\mbox{\boldmath $l$} \perp {\bf r}$
($\alpha \!=\! \pi/2$). The another type of the MH vortex is classified as $w'\!=\! -1$, 
i.e., the winding combination $\langle 2, 1, 0 \rangle$. 
As seen in Fig.~\ref{fig:2dMH}(d), the projection of the texture to the $x$-$y$ plane 
forms the cross disgyration. This texture is also called mixed-twist (MT) texture \cite{fetter,kitaPRL2001}.
These two types $\langle 0,1,2 \rangle$ and $\langle 2,1,0 \rangle$
of the coreless vortex completely degenerate at $M=0$.

\subsection{Singular vortices}

\begin{figure}[t]
\includegraphics[width=8.0cm]{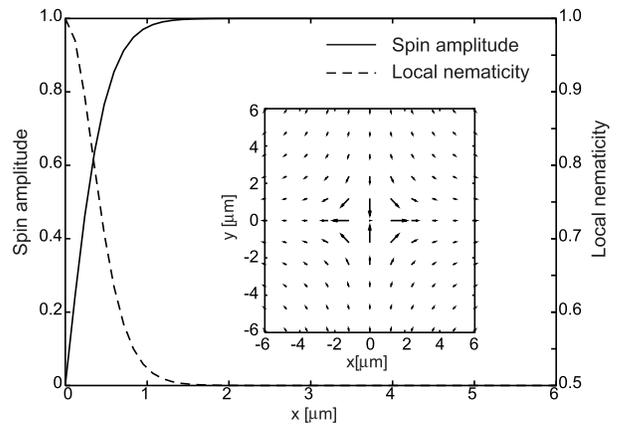} 
\caption{
The spin amplitude of the $l$-vector (solid line) and local nematicity ${\mathcal N}({\bf r})$ (dashed line)
of the polar-core vortex. The spin texture projected to the $x$-$y$ plane is shown in the inset.}
\label{fig:polar}
\end{figure}

In the high magnetization region, the external magnetic field gives the free energy dominate contribution
beyond the spin interaction energy. The texture then depends sensitively on the applied magnetic field.
The stable vortex in slowly rotating BEC's is the polar-core vortex, 
which takes the winding $w=0$ and $w'= -1$, corresponding to the combination $\langle 1, 0, -1 \rangle$. 
This vortex is thermodynamically favored over two type coreless vortices in the high magnetization region,
which is independent on the strength and the sign of the spin-interaction,
i.e., the polar-core vortex is the stable object 
for both the ferromagnetic and antiferromagnetic interaction \cite{mizushimaPRA}.
In this configuration, $\psi_0$ component with zero winding number $w_0\!=\!0$ 
occupies the central region of the vortex core which is made up by the $\psi_1$ component. 
Figure \ref{fig:polar} shows the spin amplitude and the local nematicity. 
This state is classified as the winding number $w'=-1$ 
and thus 2-dimensional texture forms the cross disgyration, 
shown in the inset of Fig.~\ref{fig:polar}.
However, this texture differs from that of the MT vortex in the followings;
(i) The polar-core vortex has the singularity at the center of the cross disgyration where the 
local nematic order grows up. (ii) Since the spin $F_z = 0$ component is localized in the core region
whose length scale is the order of the density variation 
characterized by $\xi _n=\hbar/\sqrt{2mn _0 g_n} \sim 0.1 {\rm \mu m}$,
the spatial variation of local spin and nematic orders is also the order of $\xi _n$. 
Here, $n_0$ denotes the peak density of the condensate.
As the total magnetization $M$ increases, this state continuously connects to 
the conventional vortex state in a scalar BEC. 

Recently, Leanhardt {\it et al.} \cite{leanhardt03L} have created the coreless vortex 
by using the Berry phase engineering method \cite{nakahara}.
Bulgakov and Sadreev have also found that the polar-core vortex, $\langle 1, 0, -1 \rangle$, 
becomes stable in a non-rotating harmonic trap with an applied Ioffe-Pritchard magnetic field \cite{bulgakov}.

\section{Vortex Phase diagram}

\begin{figure}[b]
\includegraphics[width=8cm]{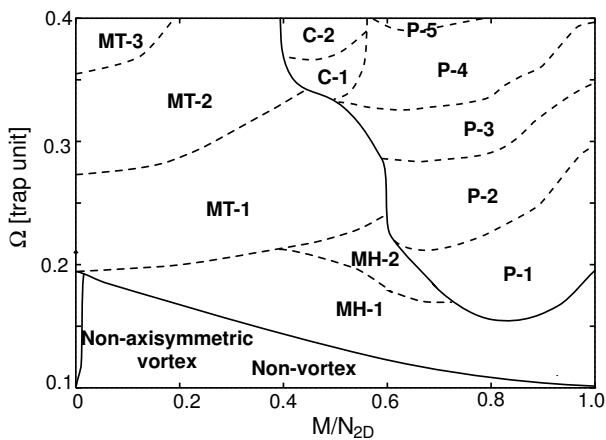}
\caption{ The vortex phase diagram in the plane by the total magnetization $M$
and the external rotation frequency $\Omega$.
The solid lines represent phase boundaries between different vortex states: the non-vortex state,
singular vortices, and continuous vortices. The details of each phase are noted in the text.}
\label{fig:phase}
\end{figure}

The vortex phase diagram is calculated by comparing the free energy 
in the plane of the total magnetization $M$ and the external rotation frequency $\Omega$.
In Fig.~\ref{fig:phase}, we display the resulting phase diagram of the vortex state
in the $M$-$\Omega$ plane up to $\Omega=0.4\omega$.
The phase diagram for slow rotation up to $\Omega=0.2\omega$ qualitatively agrees with 
the earlier result in Ref.~\cite{mizushimaPRL} 
where a few parameter is different from the current system.
The rest region of the phase diagram can be divided into two categories: 
(i) coreless vortex lattices formed by the MH and/or MT textures,
and (ii) singular vortex lattices filling the core with the polar state.
The phase boundary is depicted by the solid line in Fig.~\ref{fig:phase}.
The phase of continuous vortices can be classified as follows:
(MH-1) the single MH vortex state as seen in Figs.~\ref{fig:2dMH}(a)-(c),
(MH-2) a pair state of the off-centered MH vortex with the radial or circular disgyration,
and (MT-$n$) the vortex states of $n$ MT vortices mixing with some MH vortices ($n=1,2,3$).
Here we have identified the stable phase within the number of the MT vortex rather than 
MH vortex, because in the nonzero $M$ region the degeneracy of 
MH and MT vortices is lifted by the presence of the external magnetic field and 
the role of the MT vortex becomes important. The details will be discussed in Sec. IV A.
The other phases labeled by (P-$n$) in Fig.~\ref{fig:phase} is identified as 
the singular vortex lattices formed by a number of the polar-core vortices.
It is noted that at the limit of a scalar condensate, $M/N=1$, 
the region $\Omega < 0.2\omega$ corresponds to the vortex-free state and 
at the rotation rate $\Omega = 0.4\omega$ the 4 vortices is energetically stable.
The composite state of MT and polar-core vortices also appear in the narrow $M$ region, 
labeled by (C-1) and (C-2).

For slowly rotating BEC, $\Omega \sim 0.2 \omega$, 
two axisymmetric vortices shown in Figs.~\ref{fig:2dMH} and \ref{fig:polar} are energetically favored.
It is seen that the single MH vortex is stable in the low magnetization region of Fig.~\ref{fig:phase}
while the single polar-core vortex appears in the high magnetization region.
The separation of this stable region can be qualitatively understood from 
the energy of the external rotation drive $-\Omega L_z$.
It is easy to calculate the total angular momentum $L_z$ of axisymmetric vortices;
by using the total number $N_{\rm 2D}$ and the total magnetization $M$, it is simply written as 
\begin{eqnarray}
\frac{L_z}{\hbar N_{\rm 2D}} \!=\! w - w' \frac{M}{N_{\rm 2D}}.
\end{eqnarray}
At $M/N_{\rm 2D} = 1$, MH vortices with $w = w'=1$ are equivalent to 
the vortex-free state with $L_z/\hbar N_{\rm 2D} = 0$. 
As $M$ decreases, the angular momentum increases as a linear function of $M$
and it reaches to the value of the single vortex $L_z/\hbar N_{\rm 2D}=1$ at $M/N_{\rm 2D}=1$.
Conversely, the polar-core vortex with $w = 0$ and $w' = -1$ forms the conventional vortex 
with $L_z/\hbar N_{\rm 2D} = 1$ at $M/N_{\rm 2D}=1$ and as $M$ decreases, 
the angular momentum reaches to the zero because of the growth of the spin components 
with the zero and negative winding numbers. This contrast behavior for the total magnetization 
may be reflected by the competition of two axisymmetric vortices in the phase diagram. 
For the higher rotation frequencies beyond the single vortex region, it is also seen that 
the stable region of singular and coreless vortices are similarly separated along the magnetization.
The detailed structures of these two phases are discussed separately below.

\subsection{Coreless vortices}

In the low magnetization region, corresponding to the zero or weak magnetic field, 
the energy of the spin interaction is dominated over the magnetization term.
Here, the important length scale is characterized by the spin-interaction strength,
$\xi _s\!=\! \hbar/\sqrt{2mn_0g_s} \!\sim\! 10\xi _n$ \cite{spindomain},
which results from the competition between the spin interaction and the kinetic term.
This length scale becomes the order of spacing between neighboring vortices. 
This requires the continuous spin texture as the energetically favored state.
Such continuous vortex states can be interpreted to consist of several MH and MT vortices.

\begin{figure*}[t!]
\includegraphics[width=8cm]{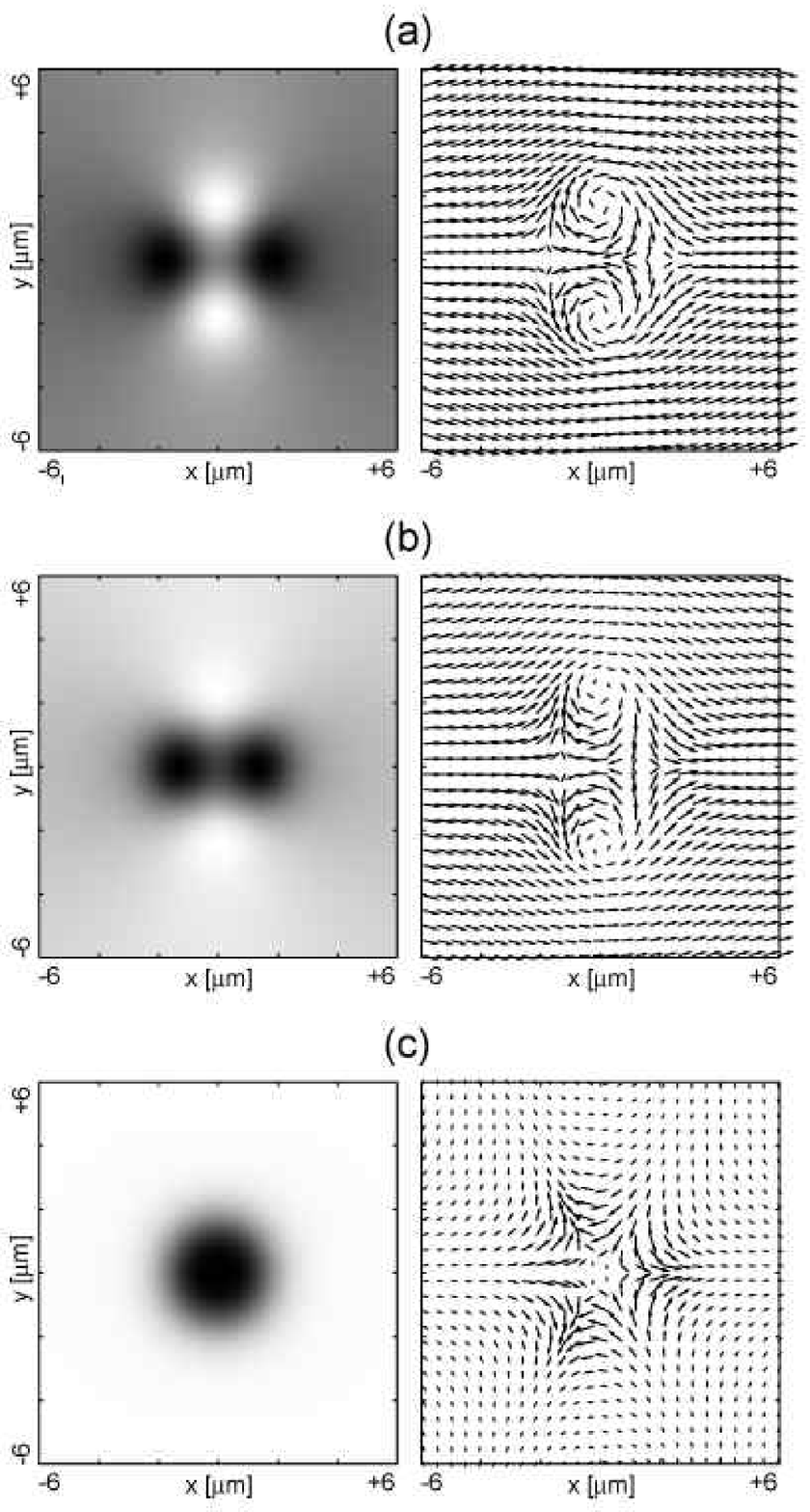} 
\includegraphics[width=8cm]{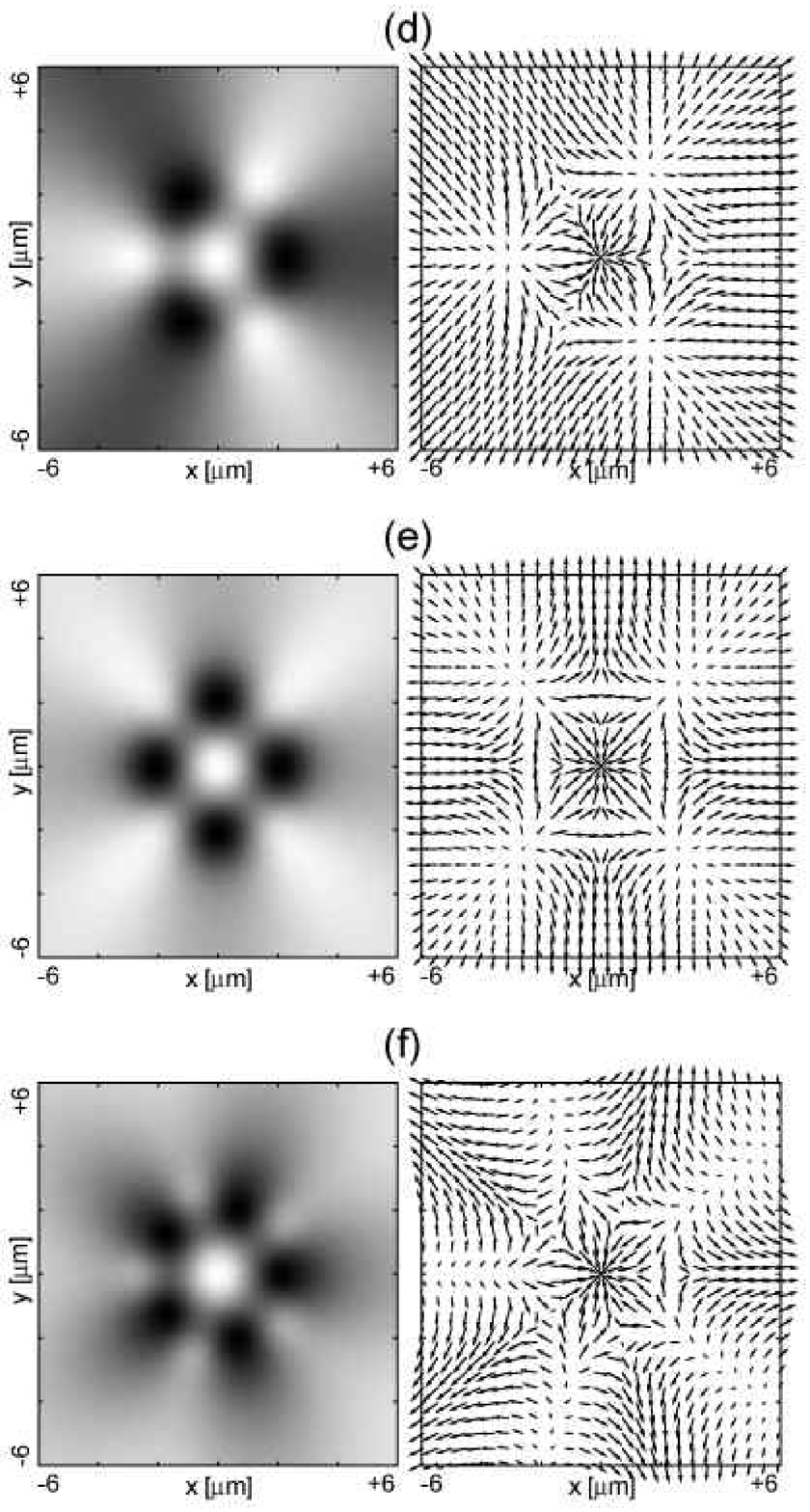} 
\caption{The textures of continuous vortex lattices near the low magnetization at $\Omega = 0.35\omega$:
the vortex states with two MH and two MT textures at $M/N=0.0$ and $L_z /\hbar N = 2.44$ (a),
$M/N=0.18$ and $L_z/\hbar N =2.47$ (b), and at $M/N=0.3$ and $L_z/\hbar N = 2.58$ (c).
The density map of $l_z$ is displayed in the left column of (a) - (c) 
in the range of $-1$ (black) and $+1$ (white).
The vector plot corresponds to the projection of the $l$-vector to the $x$-$y$ plane.
(d)-(f) are the other possible stationary state at $M/N_{\rm 2D}=0$;
$\Omega=0.4\omega$ (d), and $0.45\omega$ (e) and (f). 
All the vortex states have the continuous spin textures 
and the total density profiles always have the smooth bell shape.}
\label{fig:manyMH}
\end{figure*}

Figure \ref{fig:manyMH}(a) shows one example of the stable vortices 
at $\Omega = 0.35 \omega$ and $M/N_{\rm 2D} = 0$ where
the vector plots are the projection of the local spin $\mbox{\boldmath $l$}$
to the $x$-$y$ plane and the density map of $l_z$ represents the projection to the $z$-axis. 
This state has the continuous spin texture formed by two MH and two MT vortices
which are arranged regularly to form a square lattice.
This spin texture is similar to that proposed
by Fujita {\it et al}. \cite{fujita} in connection to superfluid $^3$He-A phase under rotation.
The total density profile forms a quite smooth bell shape.

As $M$ increases, the angular momentum of the MH vortex continuously decreases 
as a function of $M$ while that of the MT vortex increases. 
Therefore, at the finite $M$, the MT vortex becomes the important object rather than the MH vortex. 
Figures~\ref{fig:manyMH}(a)-(c) show the equilibrium state
for the different $M$: $M/N _{\rm 2D} = 0.0$ (a), 0.18 (b), 0.3 (c).
At $M=0$, four degenerate coreless vortices forms a square lattice. 
As $M$ increases, two cores with the MT texture approach to each other.
As seen in Fig.~\ref{fig:manyMH}(b), the equilibrium state at $M/N_{\rm 2D}=0.18$ forms
the double-core of two MT vortices.
Here, two MH textures with the circular disgyration start to be locked by the external magnetic field,
which implies that the MH vortex continuously deforms into the vortex-free state.
In Fig.~\ref{fig:manyMH}(c), the equilibrium texture at $M/N_{\rm 2D}=0.3$ 
and $\Omega = 0.35\omega$ is displayed 
has the axisymmetry with the higher winding number $w = +2$ and $w'=-2$, 
 i.e., $\langle 4, 2, 0 \rangle$.
This can be also regarded as a pair state of the MT vortex with $w '=-1$.
The spin texture then can be obtained as 
$\mbox{\boldmath $l$} 
= \hat{z} \cos{\beta} + \sin{\beta}(\hat{x} \cos{2\theta} - \hat{y}\sin{2\theta})$.
Here the bending angle $\beta(r)$ varies form $\beta(R)=0$ around the condensate surface 
to $\beta(0)=\pi$ at the core.
These three states are classified as two MT vortices (MT-2) in the phase diagram, Fig.\ref{fig:phase}.

As the rotation rate $\Omega$ increases, however, the vortex states with 
various discrete rotational symmetry around the BEC center 
is favored, such as the threefold (b), fourfold (c), and fivefold symmetric states (d)
in Fig.~\ref{fig:manyMH}.
It should be emphasized that all these states have no singularities in the spin texture.

\subsection{Singular vortices}

\begin{figure*}[t]
\includegraphics[width=16cm]{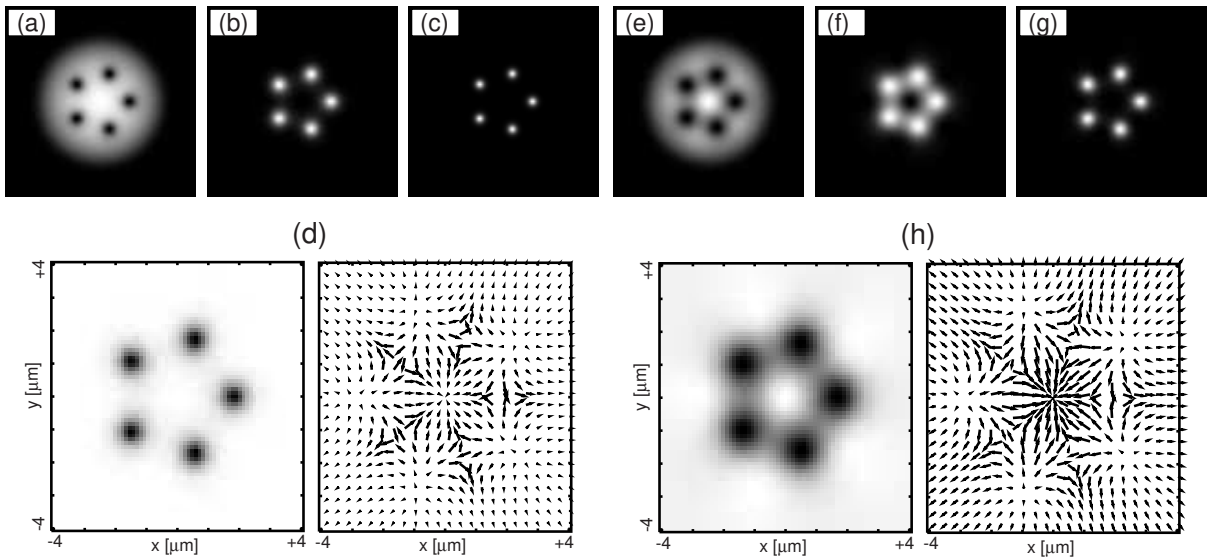}
\caption{The equilibrium state at $\Omega=0.4\omega$ and $M/N_{\rm 2D}=0.92$:
The density profiles of the spin $F_z=+1$ (a) and $F_z=0$ (b) components,
the local nematicity $\mathcal{N}$ (c) in the region 
$-6 \mbox{$\mu$m}<x,y < +6 \mbox{$\mu$m}$.
Figure (d) shows the $l$-vector in the central region $-4 \mbox{$\mu$m}<x,y < +4 \mbox{$\mu$m}$.
In the left figure, the density map of $l_z$ is shown in the range of 0 (black) and $+1$ (white),
and the vector plot in the right one corresponds to the projection of the $l$-vector to the $x$-$y$ plane.
(e)-(h) correspond to the equilibrium state at $\Omega=0.4\omega$ and $M/N_{\rm 2D}=0.71$.
The local nematicity in (c) and (g) is displayed in the range of 0.5 (black) and 1 (white).
}
\label{fig:polars}
\end{figure*}

As shown in Fig.~\ref{fig:phase}, the singular vortices labeled by P-$n$ are energetically favored 
over coreless vortices in the high magnetization region $M/N_{\rm 2D}\!>\!0.5$.
At the high magnetization limit $M/N_{\rm 2D}\!=\!1$, all spins are polarized along the $z$-axis. 
This spin-polarized gas is then described within the scalar order parameter and 
the rotating ground state forms the conventional vortex-lattice with the $m$-fold 
discrete rotational symmetry ($m\!=\!2, 3, 4, 5,$ and 6) \cite{butts}.
As $M$ decreases, the vortex core formed by the spin $F_z\!=\!+1$ component are filled by 
the other condensate with the spin $F_z\!=\!0$.
We depict the density profiles of each condensate with spin $F_{z}\!=\!+1$ and 0 components
at $\Omega = 0.4 \omega$ and $M/N_{\rm 2D} = 0.92$ in Figs.~\ref{fig:polars}(a) and (b).
This state consists of five polar-core vortices, labeled by P-5 in Fig.~\ref{fig:phase}.
The spin $F_z=0$ component is then localized in the narrow region which is the order of the length scale
of the density variation $\xi _n$.
Therefore the sharp peak of the local nematicity appears in the vortex core region, 
shown in Fig.~\ref{fig:polars}(c).
As seen in Fig.~\ref{fig:polars}(d), the almost spins are locked by the external magnetic field
and the spin texture can continuously vary around the cores.
This forms the cross disgyration having the singularities at the center of the disgyration.
The length scale is characterized by $\xi _n \ll \xi _s$ and thus the spatial variation of 
the spin texture is much shorter than the spacing between vortices. 
This results in that there is no correlation between each polar-core vortex.

Figures \ref{fig:polars}(e) and (f) show the density profiles of each component, $\rho _{+1}$ 
and $\rho _{0}$, at $\Omega = 0.4\omega$ and $M/N_{\rm 2D}=0.71$. 
As $M$ decreases, the spin 0 component $\rho _{0}$ spreads in 
between vortices formed by the spin $F_z\!=\!+1$ component.
This leads to the growth of the local nematicity, shown in Fig.~\ref{fig:polars}(g).
The spin texture is depicted in Fig.~\ref{fig:polars}(h) 
where the spin texture forms five cross disgyrations having the broader spatial variation 
than that in Fig.~\ref{fig:polars}(d).
Then, the coreless vortex with the radial disgyration is spontaneously 
created in the center of five polar-core vortices
in order to smoothly connect the five cross disgyrations.
This is similar to the texture of the coreless vortices shown in Fig.~\ref{fig:manyMH} (f).
However the length scale of the spatial variation of the texture is characterized by
the order of the density variation $\xi _n$ rather than $\xi _s$.
In addition, the point singularities exist in the spin texture.
It should be noted that the system keeps the discrete rotational symmetry 
even if the total magnetization $M$ further decreases, 
which is the common feature to the other ground states
labeled by P-$n$ in Fig.~\ref{fig:phase} ($n=2, 3,$ and 4).

\section{Vortex Lattices}

\begin{figure}[t!]
\includegraphics[width=7cm]{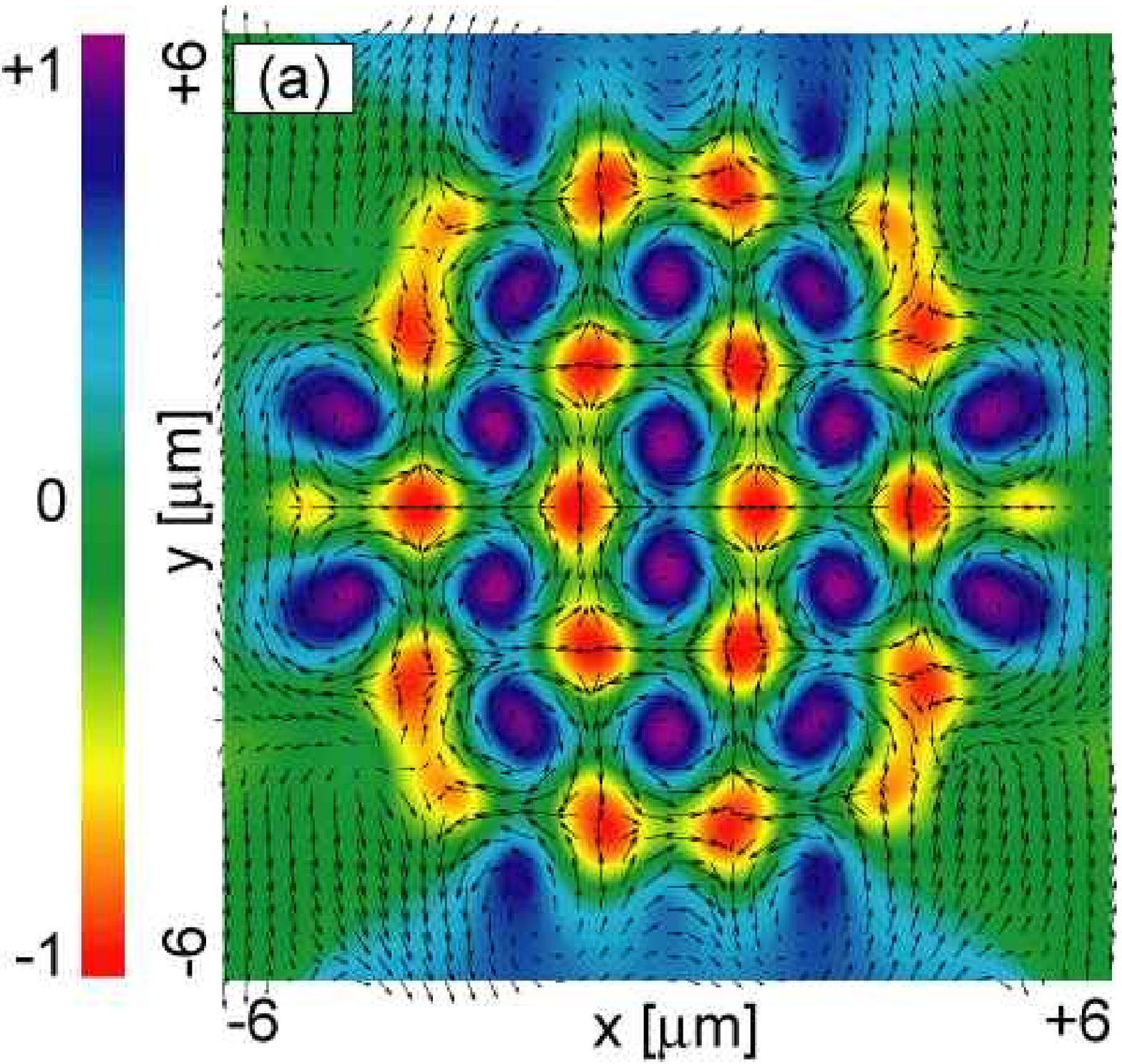} \\
\includegraphics[width=7cm]{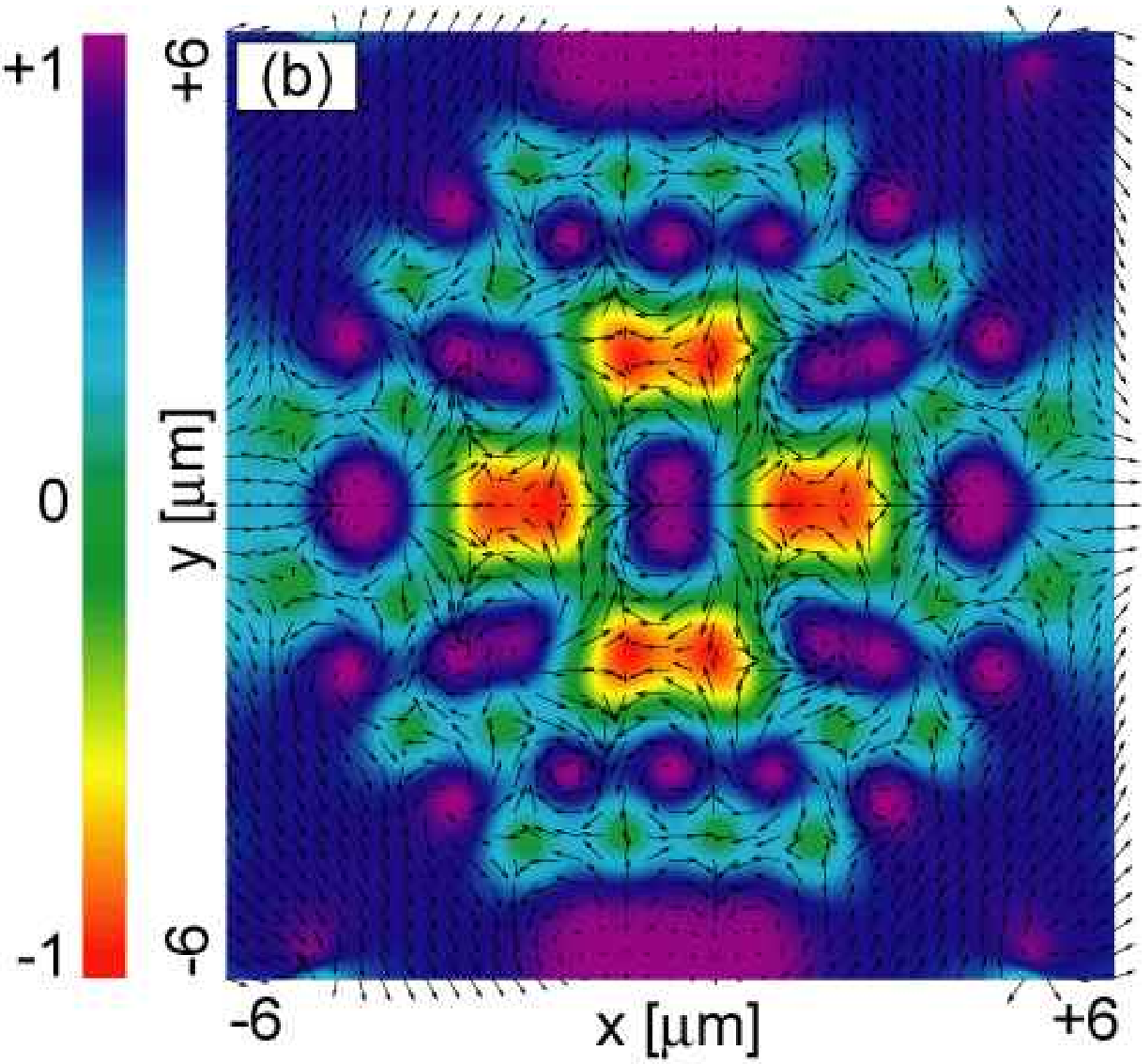} \\
\includegraphics[width=7cm]{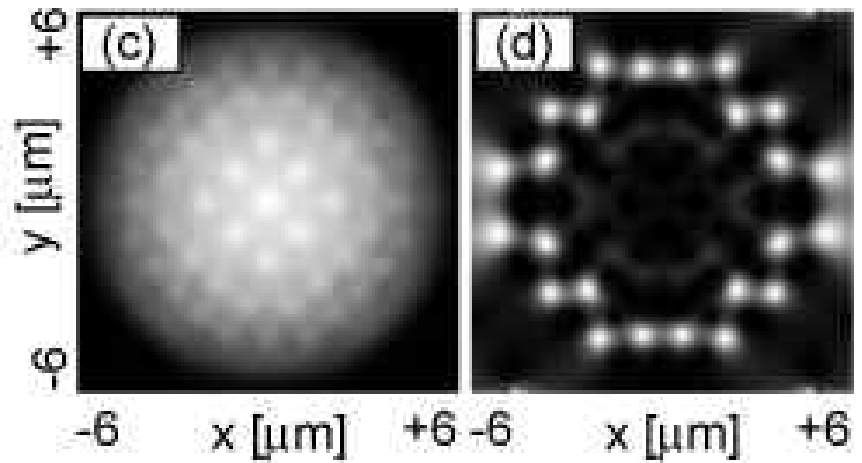} 
\caption{
The spin textures at $\Omega =0.9\omega$ and $M/N_{\rm 2D}=0.0$ (a)
and $M/N_{\rm 2D}=0.3$ (b). At  $M/N_{\rm 2D}=0.3$
the total density profile and the local nematicity are shown in (c) and (d), respectively.
The local nematicity in (d) is displayed in the range of 0.5 (black) and 1 (white).}
\label{fig:mlattice}
\end{figure}

\begin{figure}[t!]
\includegraphics[width=8cm]{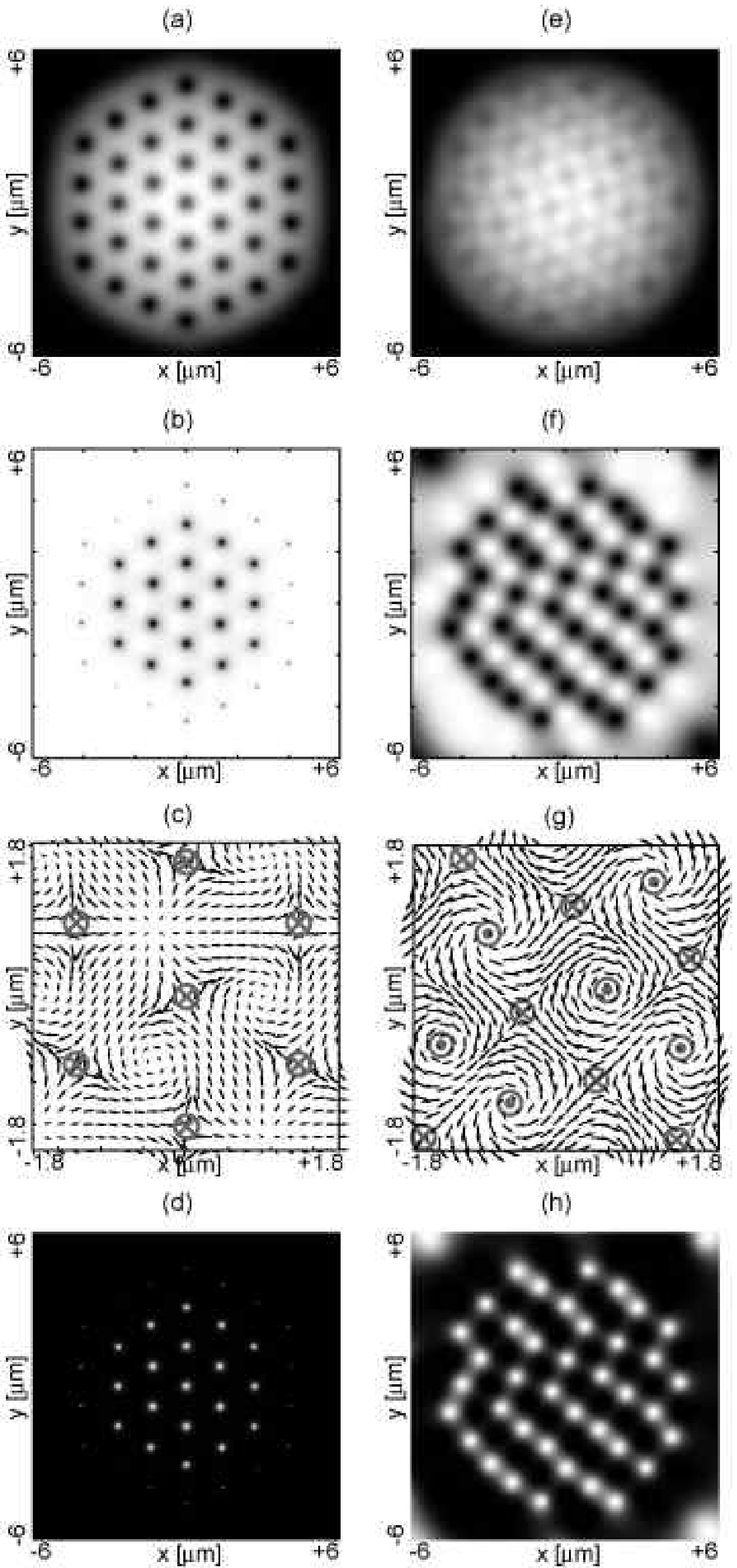} 
\caption{At $\Omega=0.9\omega$ and $M/N_{\rm 2D}=0.95$,
the total density profiles (a) and spin texture projected to the $z$-axis (b).
In (b), the density map of $l_z$ is shown in the range of 0 (black) and $+1$ (white),
The $l$-vector in the central region $-1.8\mu{\rm m} \le x, y\le +1.8 \mu {\rm m}$
is depicted in (c), where $\otimes$ means the location of and the spin singularity 
$|\mbox{\boldmath $l$}| = 0$. 
(d) The 2-dimensional profile of the local nematicity.
(e)-(h) The equilibrium state at $\Omega=0.9\omega$ and $M/N_{\rm 2D}= 0.59$.
The local nematicity in (d) and (h) is displayed in the range of 0.5 (black) and 1 (white).
}
\label{fig:plattice}
\end{figure}

Here we show one example of the equilibrium state for rapidly rotating BEC's.
Under slow rotation, the ground state at $M/N_{\rm 2D}\sim 0$ 
consists of coreless MH and/or MT vortices as seen in Sec. IV.
The initial state is taken as the continuous texture periodically arranged to the square lattice,
whose unit cell consists of two MH and two MT vortices as seen in Fig.~\ref{fig:manyMH}(a).
By numerically solving the GP equation at $\Omega=0.9\omega$ 
in the absence of the external magnetic field, 
the square lattice is given as the equilibrium state, shown in Fig.~\ref{fig:mlattice}(a).
This lattice is constructed from two sublattices of the MH vortex with the circular disgyration
and the MT vortex with the cross disgyration, 
and the local spins on the two sublattice sites are locked to the alternative direction 
$\hat{z}$ and $-\hat{z}$, respectively.
It is however found in the equilibrium spin
that there is slight distortion from a square array around the condensate surface where
the Thomas-Fermi radius is $\sim 5.8\mu \mbox{m}$.

In the presence of the weak magnetic field, the same initial state
yields the another equilibrium state.  
The equilibrium texture and the total density profile at $M/N_{\rm 2D}=0.3$ 
is depicted in Fig.~\ref{fig:mlattice}(b) and (c), respectively. 
At the BEC center, there is the non-singular skyrmion texture with the higher winding number
whose spin texture is characterized by Eq.~(\ref{eq:spin}) with $w=+2$ and $w'=+2$.
This is also identified as a pair of the MH vortex with the circular disgyration.
In addition, two MT vortices with the cross disgyration form the double-core state 
and are located around the center.
It is however found that the other vortices with the cross disgyration appear around 
the condensate surface. 
They have point singularities and thus the local nematicity 
shown in Fig.~\ref{fig:mlattice}(d) grows up at the cores. 
This vortex lattice can be identified as the composite lattice of the coreless vortex and 
the polar-core vortex. It is noted that two length scales of the spin variation 
exist in this lattice sate, $\xi _n$ and $\xi _s$.

On the other hand, we discuss on the stability of the vortex lattice 
in the high $M$ region for rapidly rotating BEC's. 
Figures \ref{fig:plattice}(a)-(d) show the structure of the equilibrium state at $\Omega = 0.9\omega$
and $M/N_{\rm 2D}=0.95$: the total density profiles (a),   
the spin texture (b) and (c), and the local nematicity (d).
The initial state in the calculation of the GP equation is taken as the vortex configuration 
with the hexagonal symmetry which is the most favored configuration in rotating scalar BEC's, 
corresponding to the high magnetization limit $M/N\!=\!1$. 
The resulting equilibrium state consists of the vortices filling the cores with the polar state and 
keeps the hexagonal symmetry.
As shown in Fig.~\ref{fig:plattice}(d), the local nematic order grows up in the narrow core region
where the spin texture has singularities. This length scale is characterized by the order 
of the density variation $\xi _n$.

Figures \ref{fig:plattice}(e) shows the total density profiles 
at $\Omega=0.9\omega$ and $M/N_{\rm 2D}=0.59$.
It is seen that the symmetry of the lattice drastically changes from the hexagonal symmetry to 
square one, where the initial state is taken as the vortex lattice 
regularly arranged to the hexagonal symmetry. 
Such the transition occurs at the total magnetization 
$M/N_{\rm 2D}\sim 0.93$ at $\Omega=0.9\omega$.
The spin texture projected to the $z$-axis is depicted in Fig.~\ref{fig:plattice}(f) and (g).
As seen in Fig.~\ref{fig:plattice}(g), the coreless MH vortex with the circular disgyration is
spontaneously created in spacing between polar-core vortices with the cross disgyration.
In addition, the spatial distribution of the local nematicity spreads over spacing 
between neighboring vortices, shown in Fig.~\ref{fig:plattice}(h).
These result from the spatial continuity of the spin texture, i.e., the ferromagnetic feature
of the spin interaction.
Such the drastic change of the symmetry 
of the vortex lattice has been also reported in two-component BEC's \cite{muellerPRL,kasamatsu}.

\section{Conclusions}

In this paper, we have presented the results of numerical calculations of 
the vortex-lattice states in rotating BEC's with the ferromagnetic interaction.
In the absence or presence of the external magnetic field, 
we have studied the local properties of the stable states, 
such as the spin texture and the local nematicity.
In addition, the vortex phase diagram has been constructed in the plane 
by the external rotation frequency $\Omega$ and the total magnetization $M$.
The stable phase can be divided into two categories: (i) coreless vortices with the non-singular spin texture
and (ii) singular lattices with polar (antiferromagnetic) cores.
In the resulting phase diagram, the competition between these two phases has been found.

Coreless lattices which are the most favored state in the zero magnetic field are formed by 
several types of the 2-dimensional disgyration, corresponding to 2-dimensional skyrmion lattices. 
In addition, for rapidly rotating BEC's, the coreless vortex regularly arranged to the square lattice 
has been presented and the stability has been discussed.
This square lattice becomes unstable for applying the weak magnetic field 
and then, neighboring MT vortices are paired and form the double-core lattice or 
the skyrmion lattice with the higher winding number.

In the high $M$ region, it has been also demonstrated that the singular lattices
are energetically favored over coreless vortices.
The projection of the texture to $x$-$y$ plane in the singular vortex 
is the same as that in the coreless MT vortex.
With decreasing $M$, the spatial variation of the spin texture becomes broader than the order 
of the density variation $\xi _n$ and the correlation between neighboring vortices plays an 
essential role. As the result, the drastic change of the hexagonal lattice into the square one
has been demonstrated, which may be experimentally identified by using the spin imaging 
within polarized light proposed by Carusotto and Mueller \cite{carusotto}.

\acknowledgments

One of the authors (T.M.) would like to acknowledge
the financial support of Japan Society for the Promotion of Science for Young Scientists.
Some numerical computation in this work has been done using the facilities of the Supercomputer Center, 
Institute for Solid State Physics, University of Tokyo.



\begin{thebibliography}{2}

\bibitem{nist}
M.R. Matthews, B.P. Anderson, P.C. Haljan, D.S. Hall, C.E. Wieman, and E.A. Cornell,
Phys. Rev. Lett. {\bf 83}, 2498 (1999).

\bibitem{ens}
K.W. Madison, F. Chevy, W. Wohlleben, and J. Dalibard, Phys. Rev. Lett. {\bf 84}, 806 (2000).

\bibitem{jila}
P.C. Haljan, I. Coddington, P. Engels, and E.A. Cornell, Phys. Rev. Lett. {\bf 87}, 210403 (2001). 

\bibitem{leanhardt}
A.E. Leanhardt, A. G\"orlitz, A.P. Chikkatur, D. Kielpinski, Y. Shin, 
D.E. Pritchard, and W. Ketterle, Phys. Rev. Lett. {\bf 89}, 190403 (2002).

\bibitem{vortex}
See for a review of vortices, A.L. Fetter and A.A. Svidzinsky, 
J. Phys.: Condens. Matter {\bf 13}, R135 (2001). 

\bibitem{abo}
J.R. Abo-Shaeer, C. Raman, J.M. Vogels, and W. Ketterle, Science {\bf 292}, 476 (2001).

\bibitem{coddington}
I. Coddington, P. Engels, V. Schweikhard, and E.A. Cornell, Phys. Rev. Lett. {\bf 91}, 100402 (2003).

\bibitem{stenger}
J. Stenger, S. Inouye, D.M. Stamper-Kurn, H.-J. Miesner, A.P. Chikkatur and W. Ketterle, 
Nature {\bf 369}, 345 (1998).

\bibitem{gorlitz}
A. G\"{o}rlitz, T. L. Gustavson, A. E. Leanhardt, R. L\"{o}w, A. P. Chikkatur, 
S. Gupta, S. Inouye, D. E. Pritchard, and W. Ketterle,  Phys. Rev. Lett. {\bf 90}, 090401 (2003).

\bibitem{barrett}
M. Barrett, J. Sauer and M.S. Chapman, Phys. Rev. Lett. {\bf 87}, 010404 (2001).

\bibitem{schmaljohann}
H. Schmaljohann, M. Erhard, J. Kronj\"{a}ger, M. Kottke, S. van Staa, L. Cacciapuoti,
J. J. Arlt, K. Bongs, and K. Sengstock,
Phys. Rev. Lett. {\bf 92}, 040402 (2004).

\bibitem{chang}
M.-S. Chang,  C.D. Hamley,  M.D. Barrett,  J.A. Sauer,  K.M. Fortier,  W. Zhang,  L. You,  
and M.S. Chapman, Phys. Rev. Lett. {\bf 92}, 140403 (2004).

\bibitem{takasu}
Y. Takasu, K. Maki, K. Komori, T. Takano, K. Honda, M. Kumakura, T. Yabuzaki, and Y. Takahashi,
Phys. Rev. Lett. {\bf 91}, 040404 (2003). 

\bibitem{ohmi}
T. Ohmi and K. Machida, J. Phys. Soc. Jpn. {\bf 67}, 1822 (1998).

\bibitem{ho}
T.-L. Ho, Phys. Rev. Lett. {\bf 81}, 742 (1998).

\bibitem{klausen}
N.N. Klausen, J.L. Bohn, and C.H. Greene, Phys. Rev. A {\bf 64}, 053602 (2001).

\bibitem{stoof}
H.T.C. Stoof, E. Vliegen, and U. Al Khawaja, Phys. Rev. Lett., {\bf 87}, 120407 (2001).

\bibitem{martikainenPRL}
J.-P. Martikainen, A. Collin, and K.-A. Suominen, Phys. Rev. Lett. {\bf 88}, 090404 (2002).

\bibitem{savage}
C. M. Savage and J. Ruostekoski, Phys. Rev. A {\bf 68}, 043604 (2003).

\bibitem{khawaja}
U. Al Khawaja and H.T.C. Stoof, Nature (London) {\bf 411}, 918 (2001);
Phys. Rev. A {\bf 64}, 043612 (2001).

\bibitem{marzlin}
K.-P. Marzlin, W. Zhang, and B.C. Sanders, Phys. Rev. A {\bf 62}, 013602 (2000).

\bibitem{tuchiya}
S. Tuchiya and S. Kurihara, J. Phys. Soc. Jpn, {\bf 70}, 1182 (2001).

\bibitem{mizushimaPRL}
T. Mizushima, K. Machida, and T. Kita, Phys. Rev. Lett. {\bf 89}, 030401 (2002).

\bibitem{reijndersPRA}
J.W. Reijnders, F.J.M. van Lankvelt,  K. Schoutens, and N. Read, 
Phys. Rev. A {\bf 69}, 023612 (2004).

\bibitem{leonhardt}
U. Leonhardt and G.E. Volovik, JETP Lett. {\bf 72}, 46 (2000).

\bibitem{zhou}
F. Zhou, Phys. Rev. Lett. {\bf 87}, 080401 (2001); 
Int. Jour. Mod. Phys. B {\bf 17}, 2643 (2003).

\bibitem{ruostekoski}
J. Ruostekoski and J.R. Anglin, Phys. Rev. Lett. {\bf 91}, 190402 (2003).

\bibitem{yip}
S.-K. Yip, Phys. Rev. Lett. {\bf 83}, 4677 (1999).

\bibitem{isoshimaJPS}
T. Isoshima, K. Machida, T. Ohmi, J. Phys. Soc. Jpn. {\bf 70}, 1604 (2001).

\bibitem{isoshimaPRA}
T. Isoshima and K. Machida, Phys. Rev. A {\bf 66}, 023602 (2002).

\bibitem{mizushimaPRA}
T. Mizushima, K. Machida, and T. Kita, Phys. Rev. A {\bf 66}, 053610 (2002).

\bibitem{martikainen}
J.-P. Martikainen, A. Collin, and K.-A. Suominen, Phys. Rev. A {\bf 66}, 053604 (2002).

\bibitem{kita}
T. Kita, T. Mizushima, and K. Machida, Phys. Rev. A {\bf 66}, 061601 (2002).

\bibitem{reijndersPRL}
J.W. Reijnders, F.J.M. van Lankvelt,  K. Schoutens, and N. Read, 
Phys. Rev. Lett. {\bf 89}, 120401 (2002); 
Phys. Rev. A {\bf 69}, 023612 (2004).

\bibitem{mueller}
E.J. Mueller, Phys. Rev. A {\bf 69}, 033606 (2004).

\bibitem{rajaraman}
R. Rajaraman, {\it Solitons and Instanton} (North-Holland, Amsterdam, 1982). 

\bibitem{sondhi}
S.L. Sondhi, A. Karlhede, S.A. Kivelson, and E.H. Rezayi, Phys. Rev. B {\bf 47}, 16419 (1993).

\bibitem{salomaa}
M.M. Salomaa and G.E. Volovik, Rev. Mod. Phys. {\bf 59}, 533 (1987).

\bibitem{bogdanov}
A.N. Bogdanov, and U.K. R\"{o}\ss ler, and A.A. Shestakov, Phys. Rev. E {\bf 67}, 016602 (2003).

\bibitem{morinari}
T. Morinari, Phys. Rev. B {\bf 65}, 064513 (2002) and references therein.

\bibitem{fujita}
T. Fujita, M. Nakahara, T. Ohmi, and T. Tsuneto, Prog. Theor, Phys. {\bf 60}, 671 (1978).

\bibitem{fetter}
A.L. Fetter, J.A. Sauls, and D.L. Stein, Phys. Rev. B {\bf 28}, 5061 (1983).

\bibitem{karimaki}
J.M. Karim\"{a}ki and E.V. Thuneberg, Phys. Rev. B {\bf 60}, 15290 (1999).

\bibitem{kitaPRL2001}
T. Kita, Phys. Rev. Lett. {\bf 86}, 834, (2001).

\bibitem{mermin}
N.D. Mermin and T.-L. Ho, Phys. Rev. Lett. {\bf 36}, 594 (1976).

\bibitem{anderson}
P.W. Anderson and G. Toulouse, Phys. Rev. Lett. {\bf 38},  508 (1977).

\bibitem{brey}
L. Brey, H.A. Fertig, R. C\^{o}t\'{e}, and A.H. MacDonald, Phys. Rev. Lett. {\bf 75}, 2562 (1995).

\bibitem{green}
A.G. Green, I.I. Kogan, and A.M. Tsvelik, Phys. Rev. B {\bf 54}, 16838 (1996).

\bibitem{carusotto}
I. Carusotto and E.J. Mueller, J. Phys. B {\bf 37}, S115 (2004).

\bibitem{leanhardt03L}
A.E. Leanhardt, Y. Shin, D. Kielpinski, 
D.E. Pritchard, and W. Ketterle, Phys. Rev. Lett. {\bf 90}, 140403 (2003).

\bibitem{nakahara}
M. Nakahara, T. Isoshima, K. Machida, S. Ogawa and T. Ohmi, Physica B {\bf 284-288}, 17 (2000);
T. Isoshima, M. Nakahara, T. Ohmi and K. Machida, Phys. Rev. A {\bf 61}, 063610 (2000)

\bibitem{bulgakov}
E.N. Bulgakov and A.F. Sadreev, Phys. Rev. Lett. {\bf 90}, 200401 (2003).

\bibitem{spindomain}
T. Isoshima, K. Machida, and T. Ohmi, Phys. Rev. A {\bf 60}, 4857 (1999).

\bibitem{butts}
D.A. Butts and D.S. Rokhsar, Nature (London) {\bf 397}, 327 (1999).

\bibitem{muellerPRL}
E.J. Mueller and T.-L Ho, Phys. Rev. Lett. {\bf 88}, 180403 (2002).

\bibitem{kasamatsu}
K. Kasamatsu, M. Tsubota, and M. Ueda, Phys. Rev. Lett. {\bf 91}, 150406 (2003).


\end{thebibliography}
\end{document}